\documentclass{ifacconf}

\usepackage{graphicx}      
\usepackage{natbib}        
\usepackage{amsfonts}
\usepackage{amsmath}
\usepackage{tikz}

\newcommand{\eqdef}{\overset{def}{=}}
\newtheorem{define}{Definition}
\newtheorem{problem}{Problem}
\newtheorem{remark}{Remark}

\begin{document}
\begin{frontmatter}

\title{Optimal Remote Estimation Over Use-Dependent Packet-Drop Channels - Extended Version }


\author[First]{David Ward} 
\author[Second]{Nuno C. Martins}

\address[First]{Department of Electrical and Computer Engineering and the Institute for Systems Research at University of Maryland, College Park, MD, 20742 USA (e-mail: dward2@ umd.edu).}
\address[Second]{Department of Electrical and Computer Engineering and the Institute for Systems Research at University of Maryland, College Park, MD, 20742 USA (e-mail: nmartins@ umd.edu)}

\begin{abstract}                
Consider a discrete-time remote estimation system formed by an encoder, a transmission policy, a channel, and a remote estimator. The encoder assesses a random process that the remote estimator seeks to estimate based on information sent to it by the encoder via the channel. The channel is affected by Bernoulli drops.  The instantaneous probability of a drop is governed by a finite state machine (FSM). The state of the FSM is denoted as the channel state. At each time step, the encoder decides whether to attempt a transmission through the packet-drop link. The sequence of encoder decisions is the input to the FSM. This paper seeks to design an encoder, transmission policy and remote estimator that minimize a finite-horizon mean squared error cost. We present two structural results. The first result in which we assume that the process to be estimated is white and Gaussian, we show that there is an optimal transmission policy governed by a threshold on the estimation error. The second result characterizes optimal symmetric transmission policies for the case when the measured process is the state of a scalar linear time-invariant plant driven by white Gaussian noise. Use-dependent packet-drop channels can be used to quantify the effect of transmission on channel quality when the encoder is powered by energy harvesting. An application to a mixed initiative system in which a human operator performs visual search tasks is also presented.
\end{abstract}


\end{frontmatter}

\section{Introduction}

Encoders often select varying channel modes to enhance transmission performance in the presence of power and energy constraints. For example, in battery-operated wireless communication systems with energy harvesting, the decision of whether to attempt transmission must be made time and again at each time-step. The charge-level of the battery induces memory in the channel, which must be monitored for use by the transmission policy. We define a class of \textit{use-dependent packet-drop channels} to model the effect of attempted transmissions on current and future performance, which in our case is quantified by the probability that an attempted transmission is dropped. The memory in use-dependent packet-drop channels is modeled by a finite state machine (FSM). The state of the FSM, or channel state, determines the instantaneous probability of drop. In our formulation the only relevant input to the FSM is the time-sequence of decisions of whether to attempt a transmission.

We consider a system formed by a remote estimator, a transmission policy, a use-dependent packet-drop channel and an encoder. The estimator produces  an estimate of the state of a linear time-invariant plant that is accessible to the encoder. The estimate is based on information transmitted from the encoder to the estimator via the channel. The encoder and transmission policy also have access to past transmission decisions and channel feedback on the realization of current and past drops. The encoder determines what to transmit over the channel and the transmission policy determines when to attempt a transmission. The main goal of this paper is to investigate encoders, transmission policies  and remote estimators that jointly minimize the mean squared  state estimation error over a finite time-horizon.  Section \ref{problemFormulation} contains the problem formulation.

\subsection{Outline of the main results}

The following are our two main results characterizing the structure of optimal transmission policies for our problem. 

In the first result, we assume that the process to be estimated is white and Gaussian. We show that the optimal transmission policy is of the threshold type, meaning that the encoder chooses to attempt transmission when the process takes values outside a certain interval $[\underline{\tau}, \bar{\tau}]$. The characteristics of the use-dependent packet-drop channel determine the values of $\bar{\tau}$ and $\underline{\tau}$. In general, $\bar{\tau}$ may not equal $-\underline{\tau}$, even when the process is zero-mean.

In the second result, the process to be estimated is the state of a scalar linear time-invariant plant driven by white Gaussian noise, for which we seek to obtain an optimal symmetric  transmission policy. We show that if the channel performs satisfactorily in all channel states, then there exists at least one symmetric threshold that, when applied to the estimation error, leads to a transmission policy that is optimal among all symmetric strategies. We present a numerical example that illustrates, for specific classes of use-dependent channels, that threshold policies are optimal among all symmetric strategies, even when there are no restrictions on the performance of the channel.

In section \ref{problemFormulation}, the formal definition of use-dependent packet-drop channels is given and the problem is formulated. \mbox{Section \ref{sec:results}} presents the technical results. Section \ref{sec:appsBig} outlines two engineering applications of our formulation. The Appendix presents basic concepts on quasi-convex functions.

\begin{figure}[t]
\centering
\tikzstyle{every node}=[font=\large]
{\begin{tikzpicture}[xscale=.8, yscale=1]

\draw (-.05,0) rectangle (1.55,1);
\node at (.8,.5) {$\mathcal{U}_n, \,{\mathcal{E}}_n$};
\node[below left] at (.7,0) {$C_n$};

\draw (3, 0) rectangle (5, 1);
\node[above] at (4,0) {$C_n$};

\draw (6.8, 0) rectangle (7.8,1);
\node at ( 7.3, .5 ) {$\mathcal{D}_n$};

\draw[dashed] (2.3, -.2) rectangle (5.9,4.05);
\path (0,3.4);

\node[below right,  align=center, font=\small] at (2.8, 4) { Use-Dependent \\  Packet Drop \\  Channel};

\draw (3.25, 1.6) rectangle (4.75, 2.4);
\node at (4, 2) {$\mathcal{M}$};

\node[left] at (-.5,.5) {$X_n$};
\draw[->, line width=.5mm] (-.5,.5) -- (-.05,.5);
\node[left] at (.75,1.75) {$R_n$};
\node[above right,  font=\footnotesize] at (-1.8, 3) {$R_n=1$ - Attempt};
\node[above right,  font=\footnotesize] at (-.2, 2.75) {transmission};
\node[above right,  font=\footnotesize] at (-1.8, 3.6){$R_n=0$ - Do not};
\node[above right,  font=\footnotesize] at (-.2, 3.35){transmit};
\draw[->, line width=.5mm] (.75,1) -- (.75,2.75) -- (2.3,2.75) ;
\draw[->, line width=.5mm] (2.3,2.75) -- (7.3,2.75) -- (7.3,1);
\draw[->, line width=.5mm] (4,2.75) -- (4,2.4);
\node[above right] at (1.5,.55 ){$Z_n$}; 	
\draw[->, line width=.5mm] (1.55,.5) -- (2.3,.5);
\draw[-, line width=.5mm] (2.3,.5) -- (3,.5);
\draw[->, line width=.5mm] (3,.5) -- (3.7,.5) -- (4.2, .7);
\draw[-, line width=.5mm] (4.2,.5) -- (5, .5);
\draw[->, line width=.5mm] (4, 1.6) -- (4, 1);
\node[right] at (4,1.3){$P_n$};
\draw[->, line width=.5mm] (5, .5) -- (6.8,.5);
\node[above left] at (6.8,.5 ){${V}_n$}; 
\draw[->, line width=.5mm] (7.8,.5) -- (8.3,.5);

\draw[dashed, ->,line width=.5mm] (4.1, 0 ) -- (4.1, -.4) -- (.75, -.4) -- (.75,0);
\node[below,  font=\footnotesize] at (2.425, -.4) {Channel Feedback};

\node[right] at (8.3,.5) {$\hat{X}_n$};

\end{tikzpicture}}
\caption{The problem under investigation is a remote estimation problem over a packet-drop channel, whose probability of drop $P_n$ is governed by the Finite State Machine $\mathcal{M}$.  }
\label{fig:GabReform}
\end{figure}
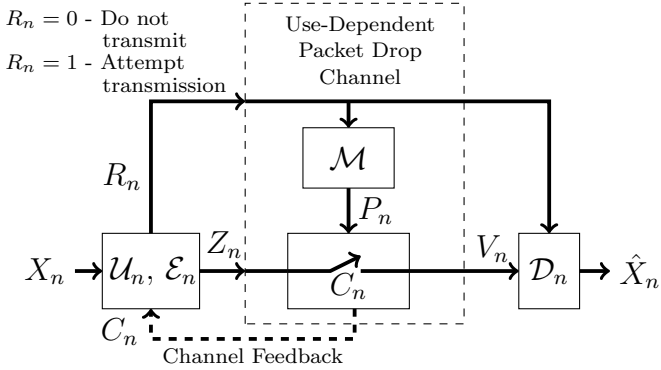

\subsection{Related Literature}

In \cite{GabrielAllerton} and  \cite{GabrielTrans}, an estimation problem over a packet drop channel with communication costs is considered. In contrast to \cite{GabrielAllerton} and  \cite{GabrielTrans}, here we introduce a channel state and do not consider explicit communication costs. In our formulation, the channel state, which depends on current and past transmission decisions, and its impact on performance create an implicit communication cost. For example, in the energy harvesting application explained in section \ref{sec:apps}, there is no explicit cost for attempting a transmission. However, attempting a transmission reduces the energy available for future transmissions, which causes performance degradation that can be viewed as an implicit cost for attempting a transmission.

Considering costly measurements (or transmissions)  in estimation and control problems has a long history and has been modeled in many ways. In \cite{Athans}, one of several possible measurements with different observation costs is selected to minimize a combination of error and observation cost. In \cite{submodular}, a subset of the measurements is selected in order to minimize the log-determinant of the error covariance. In \cite{sinopoli}, the arrival of observations is a random process and the convergence of the error covariance is studied. In \cite{paging}, the task is to locate a mobile agent and the observation cost is the expected number of observations that must be made to do so. 

In \cite{weissman2010capacity}, the capacity of channels with action-dependent states is studied. Although our problem formulation is similar to that of \cite{weissman2010capacity} in motivation, it differs in several accounts. In contrast to \cite{weissman2010capacity}, we consider finite time horizons, a mean-squared error cost and a new class of packet-drop channels. 

\section{Problem Formulation} \label{problemFormulation}

\subsection{Notation}

We use calligraphic font ($\mathcal{F}$) to denote deterministic functions, capital letters ($X$) to represent random variables and lower case letters ($x$) to represent realizations of the random variables. Let $\mathcal{N}(0, \sigma^2)$ denote the Gaussian distribution with zero mean and variance $\sigma^2$. We use $Q^n$ to denote the finite sequence $\{ Q_1,\,Q_2,\,\dots Q_n \}$. The real line is denoted with $\mathbb{R}$ and a subset of $\mathbb{R}$ is denoted with double barred font, such as $\mathbb{A}$. The indicator function of a set $\mathbb{A}$ is defined as \begin{eqnarray*}\textbf{1}_{\mathbb{A}}(x) \eqdef \begin{cases} 1 & x \in \mathbb{A} \\ 0 & \text{otherwise.}  \end{cases} \end{eqnarray*} The expectation operator is denoted with $E[\cdot]$. By $\lim_{\delta \downarrow 0} \mathcal{F}(\delta)$ we mean the limit of $\mathcal{F}(x)$ at $0$ from the right.

\subsection{Problem Formulation}

Consider the following scalar linear time-invariant system \begin{eqnarray*} X_{n+1} &=& aX_n + W_n, \quad n\geq 0, \quad X_0=x_0,\end{eqnarray*} where $X_n$ is the state, $a$ is a real constant, $W_n$ is independent and identically distributed Gaussian noise with zero mean and variance $\sigma^2$. The initial state $x_o\in\mathbb{R}$ is known.

Observations are made by the encoder and transmitted to the remote estimator over a use-dependent packet-drop channel, which is defined below.

\begin{define}[Use-dependent packet-drop channels] Let \\ $\mathcal{M}^s:\mathbb{Q}\times \{0,1\} \rightarrow \mathbb{Q}$ and $\mathcal{M}^o:\mathbb{Q}\rightarrow [0,1]$ be given, where $\mathbb{Q} = \{1,\dots \, , m \}$ represents the set of possible states of a finite state machine (FSM). The channel inputs are $Z_n$ and $R_n$, which take values in $\mathbb{R}$ and $\{ 0,1 \}$, respectively. In this model $Z_n$ represents the information to be transmitted, while the decision to attempt a transmission (or not) is represented by $R_n=1$  ($R_n=0$). The channel output $V_n$ takes values in $\mathbb{R}\cup \mathfrak{E}$ and is determined as follows 
\begin{eqnarray*}V_n &=& \begin{cases}Z_n & \text{ if }  L_n=1 \\ \mathfrak{E} & \text{ if } L_n=0, \end{cases} \end{eqnarray*} where $L_n \eqdef R_nC_n$. Here, $C_n$ is a Bernoulli process characterized by $p(C_n=0)=\mathcal{M}^o(Q_n)$, where $Q_n$ is the state of the FSM updated by \begin{eqnarray*} Q_{n+1} &=& \mathcal{M}^s(Q_n,\,R_n). \end{eqnarray*} The FSM's initial state $q_1\in\mathbb{Q}$ is known. Here, $\mathcal{M}^s$ and $\mathcal{M}^o$ model the effect of the input on the transitions among channel states and the probability of drop as a function of the channel state, respectively. 
\end{define}

In figure \ref{fig:GabReform}, the dotted box represents the use-dependent packet-drop channel. Section \ref{sec:appsBig} discusses two applications of use-dependent packet drop channels. 

At time $n$, the transmission policy \\ \mbox{${\mathcal{U}}_n: \mathbb{R}^n \times \{0,1\}^{n-1} \rightarrow \{0,1\}$} determines whether a transmission is attempted,  \begin{eqnarray*} R_n = {\mathcal{U}}_n(X^n,C^{n-1}), \end{eqnarray*}based on the plant history $X^n$ and drop history $C^{n-1}$.  The remote estimator $\mathcal{D}_n: \mathbb{R}^n \times \{0,1\}^{n} \rightarrow \mathbb{R}$ produces the state estimate, \begin{eqnarray*} \hat{X}_n = \mathcal{D}_n(V^n, R^n),\end{eqnarray*}based on the channel output history $V^n$ and the transmission history $ R^n$. The encoder $\mathcal{E}_n: \mathbb{R}^n \times \{0,1\}^{n-1} \rightarrow \mathbb{R} $ determines what is transmitted,\begin{eqnarray*}   Z_n = {\mathcal{E}}_n(X^n,C^{n-1}),\end{eqnarray*} based on the plant history $X^n$ and drop history $C^{n-1}$. 

We seek to solve the following problem. \\  \rule{\columnwidth}{1pt} \begin{problem}\label{prob:first} For finite $N$, solve \begin{eqnarray*} \underset{ \mathcal{U}^N,{\mathcal{E}}^N, \mathcal{D}^N }{\text{minimize }} \sum_{n=1}^N E\left[(X_n-\hat{X}_n)^2\right].\end{eqnarray*}   \end{problem}  \rule{\columnwidth}{1pt}

\begin{remark} \label{ProbSimplify} For any encoder and transmission policy, the optimal remote estimator is the conditional mean,  $ \mathcal{D}_n(V^n, R^n) = E[X_n|V^n, R^n]$. Also, an optimal encoder policy transmits only the current state, ${\mathcal{E}}_n(X^n,\, C^{n-1})  = X_n$. This is evident from the Markov nature of $X_n$ and the information already available to the remote estimator. The channel drops can be calculated from $(V^{n-1},R^{n-1})$; thus, the only new information to send the remote estimator is $X_n$. \end{remark}

Because of Remark \ref{ProbSimplify}, Problem \ref{prob:first} is equivalent to the following problem. \\ \rule{\columnwidth}{1pt}
 \begin{problem}[Main Problem] \label{prob:prob} For finite $N$, solve  \begin{eqnarray*} \underset{\mathcal{U}^N}{\text{minimize }}  \sum_{n=1}^N E\left[(X_n-\hat{X}_n)^2\right], \end{eqnarray*} where the optimal encoder, ${\mathcal{E}}_n(X^n,\, C^{n-1})  = X_n$, and optimal remote estimator, $\mathcal{D}_n(V^n, R^n) = E[X_n|V^n, R^n]$, are used. \end{problem} \rule{\columnwidth}{1pt}


\section{Structural results} \label{sec:results}

In this section, we present our technical results. We began by defining threshold transmission policies.

\subsection{Definitions}

Estimation error is denoted as $E_n \eqdef X_n - \hat{X}_n$.

\begin{define} A function $\mathcal{G}:\mathbb{R}\rightarrow [0,1]$ is a threshold function if there are constants $\underline{\tau}$ and$\bar{\tau}$, such that: \begin{eqnarray*} \mathcal{G}(e) = \begin{cases} 1 & \text{if } \underline{\tau} \leq e \leq \bar{\tau} \\ 0 & \text{otherwise.}\end{cases} \end{eqnarray*} \end{define}

\begin{define} A function $\mathcal{G}:\mathbb{R}\rightarrow [0,1]$ is a symmetric threshold function if there is a constant $\tau$, such that: \begin{eqnarray*} \mathcal{G}(e) = \begin{cases} 1 & \text{if } |e| \leq \tau \\ 0 & \text{otherwise.}\end{cases} \end{eqnarray*} \end{define}

\begin{define} A transmission policy $\mathcal{T}^N$ is a threshold policy if the decision to transmit depends only on the current error and channel state $(e_n, q_n)$ in the following manner \begin{eqnarray*} \mathcal{T}_n(x^n, c^{n-1}) = \begin{cases} 1 & \text{if } \underline{\tau}_n(q_n) \leq e_n \leq \bar{\tau}_n(q_n) \\ 0 & \text{Otherwise,} \end{cases} \end{eqnarray*} for some $\underline{\tau}_n(q_n)$, $\bar{\tau}_n(q_n)\in \mathbb{R} $. \end{define}

Notice that the current channel state and error $(e_n, q_n)$ are a function of the history ($x^n$, $c^{n-1}$) and previous policies $\mathcal{T}^{n-1}$.

\begin{define} A transmission policy $\mathcal{T}^N$ is a symmetric threshold policy if the decision to transmit depends only on the current error and channel state $(e_n, q_n)$ in the following manner \begin{eqnarray*} \mathcal{T}_n(x^n, c^{n-1}) = \begin{cases} 1 & \text{if } | e_n| \leq {\tau}_n(q_n) \\ 0 & \text{Otherwise,} \end{cases} \end{eqnarray*} for some ${\tau}_n(q_n) \in \mathbb{R} $. \end{define}

\subsection{Optimal transmission policies are threshold when the process is white and Gaussian ($a=0$) } \label{independent}

To investigate the structure of solutions to Problem \ref{prob:prob}, we start with the case when $a=0$. The system state becomes \begin{eqnarray*} X_n = W_n.\end{eqnarray*} Since the estimation error is independent at each step, there are optimal transmission policies that only depend on the channel state and current error. 

With $a=0$, we reformulate Problem \ref{prob:prob} as a dynamic program to show that there are optimal transmission policies of the threshold type, which may not be symmetric. An optimal transmission policy that is not symmetric in the estimation error is surprising since the cost function is symmetric in the error and the random process is zero-mean and symmetric.  

We utilize the results in \cite{MarcosAllerton}. In \cite{MarcosAllerton}, a single stage estimation problem over a  collision channel with two transmitters is studied. If both transmit then the remote estimator receives a collision symbol and if neither transmits a no-transmission symbol is received. The result in \cite{MarcosAllerton} states that the optimal policy for each transmitter is of the threshold type. 

\begin{remark} In Problem \ref{prob:prob}, when a transmission is attempted but is dropped, the remote estimator receives $(V_n= \emptyset, R_n=1)$. This is distinguishable from when no transmission is attempted $(V_n=\emptyset, R_n=0)$. In \cite{MarcosAllerton}, because the remote estimator can distinguish between a collision and a no-transmission, the optimal policies are of the threshold type and may not be symmetric. Similarly for Problem \ref{prob:prob}, the remote estimator's ability to distinguish a failed transmission and no transmission leads to optimal policies that are of the threshold type and may not be symmetric. \end{remark}

Problem \ref{prob:prob} is a sequential problem; distinguishing it from \cite{MarcosAllerton}, which is a static problem. Notice that our problem cannot be converted into a sequence of static problems because the transmission policies depend on the channel memory.

Following \cite{MarcosAllerton}, the stage cost at time $n$ can be written as \begin{align} \label{eq:stageCost} E[ (X_n-\hat{X}_n)^2 ] = &E[ (X_n-\hat{X}_n)^2 | L_n=0]p(L_n=0)& \nonumber \\ = &E[ (X_n-\hat{X}_n^0)^2 | R_n=0]p(R_n=0) &\nonumber \\  &+ p_{n} E[ (X_n-\hat{X}_n^1)^2 | R_n=1]p(R_n=1) & \end{align} where $p_n \eqdef \mathcal{M}^o (q_n)$, $ \hat{X}^0_n \eqdef E[X_n|R_n=0] $ and $  \hat{X}^1_n \eqdef E[X_n|R_n=1]  $. 

\begin{prop} \label{prop:AZero} The stage cost at time $n$ is a function of only the current channel state $Q_n$ and transmission policy $\mathcal{U}_n$. \end{prop}

\begin{pf} From (\ref{eq:stageCost}), note that $E[(X_n-\hat{X}_n)^2]$ is a deterministic function of the channel state $q_n$, the probability that $R_n=1$ and the distribution $f_{X_n|R_n}$. This distribution can be written as \begin{eqnarray*} f_{X_n|R_n}(x_n|r_n) &=& \frac{p_{R_n|X_n}(r_n|x_n)f_{X_n}(x_n)}{p_{R_n}(r_n)}, \end{eqnarray*} where $p_{R_n|X_n}(r_n|x_n)\eqdef p(R_n = r_n|X_n = x_n)$ and $p_{R_n}(r_n) \eqdef p(R_n = r_n)  $.  Thus, (\ref{eq:stageCost}) is a function of $q_n$ and the probability mass function $p_{R_n|X_n=x_n}$.  

The transmission policy $ \mathcal{U}_n$ determines the distribution $p_{R_n|X_n=x_n}$. Therefore, the stage cost is a function of only $q_n$ and $\mathcal{U}_n$. \end{pf}

With $a=0$, Problem \ref{prob:prob} can be written as a Markov chain with $\mathcal{U}_n$ as the input, $(X_n, C_n)$ as the noise, $(Q_n, X^{n-1}, C^{n-1})$ as the state, and $E[(X_n-\hat{X}_n)^2]$ as the stage cost. Note the input is not $r_n$, the decision to transmit, as may have been expected. The transmission policy  $\mathcal{U}_n$ is taken as the input because the distribution $f_{X_n|R_n}$ depends on the entire policy $\mathcal{U}_n$: not just the specific decision $r_n$.  

Using Proposition \ref{prop:AZero} and the independence of the system states over time, without loss of performance, we need to consider only transmission policies that are functions of the current system state and channel state, $ \mathcal{U}_n(X_n, Q_n)$.  Consequently, the Markov decision process can be simplified with $\mathcal{U}_n$ as the input, $X_n$ as the noise, $Q_n$ as  the state, and $E[(X_n-\hat{X}_n)^2]$ as the stage cost. The associated dynamic programming recursion is shown in (\ref{eq:Terminal}) and (\ref{eq:DP}) on the next page.

\begin{figure*}  \begin{eqnarray} \label{eq:Terminal} V_{N+1}(q_{N+1}) &=& 0 \\  \label{eq:DP} V_n(q_n) &=& \min_{\mathcal{U}_n} E[ (X_n-\hat{X}_n^0)^2 | R_n=0]p(R_n=0|q_n) + \nonumber \\ && \,\,\,\,\,\,\,p_{n} E[ (X_n-\hat{X}_n^1)^2 | R_n=1]p(R_n=1|q_n) + E[V_{n+1}(q_{n+1})| q_n], \quad n\in\{ 1, \dots, \, N \} \end{eqnarray}\begin{eqnarray} \label{big2} \frac{V_n(a(c+\delta)+w,q) - V_n(ac+w,q)}{(c+\delta)^2-c^2} &\leq& \frac{(a(c+\delta)+w)^2 - (ac+w)^2}{(c+\delta)^2-c^2} + \frac{h_{n+1}^{q^1}(c+\delta +w/a) - h_{n+1}^{q^1}(c+w/a)}{(c+\delta)^2-c^2}\\ \label{Big3} &\leq& 2a^2 +\frac{a \omega}{x} + \frac{h_{n+1}^{q^1}(c+\delta +w/a) - h_{n+1}^{q^1}(c+w/a)}{(c+\delta)^2-c^2} \end{eqnarray}  \end{figure*}

\begin{thm}\label{thm:thres} Let $X_n$ be independent and identically distributed $\mathcal{N}(0,\sigma^2)$. The optimal transmission policy for Problem \ref{prob:prob} is of the threshold type. \end{thm}

\begin{pf}For an arbitrary transmission policy $\mathcal{U}^N$, we seek a policy $\mathcal{T}^N$ that outperforms it and is a threshold policy. Note, all quantities associated with the policy $\mathcal{T}^N$ have a superscript $\mathcal{T}$.  Also, all quantities associated with policy $\mathcal{U}^N$ have a superscript $\mathcal{U}$.

We expand our search for a policy $\mathcal{T}^N$ to include randomized transmission policies. For $n \in \{1,\dots,N\}$ and $q\in \mathbb{Q}$, let $\mathcal{T}^q_n : \mathbb{R} \rightarrow [0,1]$ be the probability of transmitting, $\mathcal{T}_n^q(x) \eqdef p^\mathcal{T}(R_n=1|X_n=x,\, Q_n=q)$. Also, $ E[\mathcal{T}_n^q(X_n)] = p^\mathcal{T}(R_n=1|Q_n=q)$.

For a specific $n$, consider a policy $\mathcal{T}_n$ that matches the policy $\mathcal{U}_n$'s probability of transmitting, \begin{eqnarray}\label{eq:RnEquality} p^\mathcal{T}(R_n=1|Q_n = q_n) = p^\mathcal{U}(R_n=1|Q_n = q_n).\end{eqnarray} Also, let policy $\mathcal{T}_n$ be such that it produces estimates that match those of policy $\mathcal{U}_n$, \begin{eqnarray}\label{eq:EstimateEquality1} ^{\mathcal{T}}\hat{X}_n^0 &=& ^{\mathcal{U}}\hat{X}_n^0 \\ \label{eq:EstimateEquality2} ^{\mathcal{T}}\hat{X}_n^1 &=&  ^{\mathcal{U}}\hat{X}_n^1.  \end{eqnarray} Since $p^\mathcal{T}(R_n=1|Q_n=q)=p^\mathcal{U}(R_n=1|Q_n=q)$, we have  $p^\mathcal{T}(Q_{n+1}|Q_n = q) = p^\mathcal{U}(Q_{n+1}| Q_n=q)$. All the quantities in (\ref{eq:DP}) are the same for both policies with the exception of $E[(X_n-\hat{X}_n^i)^2|R_n=i]$, for $i=1,2$. We will choose $\mathcal{T}_n^q$ to reduce  $E[(X_n-\hat{X}_n^i)^2|R_n=i]$, for $i=1,2$. 

 In \cite{MarcosAllerton}, minimizing $E[(X_n-\hat{X}_n^i)^2|R_n=i]$ for $i=1,2$ subject to the constraints (\ref{eq:RnEquality}), (\ref{eq:EstimateEquality1})  and (\ref{eq:EstimateEquality2}) was cleverly rewritten as a constrained moment matching problem. It was shown that the optimal $\mathcal{T}_n^q$ was a threshold function of $X_n$. Using this result, we have constructed a threshold policy $\mathcal{T}_n^q$ that outperforms $\mathcal{U}_n^q$.
 
 Thus, for every $q\in \mathbb{Q}$ and $n\in\{1,\dots,N\}$, we can construct a threshold policy $\mathcal{T}_n^q$ that out forms $\mathcal{U}_n^q$. This threshold policy $\mathcal{T}^N$ outperforms $\mathcal{U}^N$.
 \end{pf}


\subsection{Optimal Solutions within the class of symmetric policies} \label{symmetric}

We now investigate the structure of the best symmetric transmission policies. We seek conditions under which the optimal symmetric transmission policy is a symmetric threshold policy. This is the case if the probability of drop is sufficiently small for all channel states. Even if the drop probabilities are not sufficiently small, symmetric threshold policies may still be optimal. This is highlighted by a numerical example, which suggests that there are classes of channel dynamics for which symmetric threshold policies are the best symmetric transmission policies. This is the topic of future research.

Restricting to symmetric transmission policies, Problem \ref{prob:prob} can be written as a dynamic program. We first show that the cost-to-go functions are quasi-convex. In order to accomplish this, we write the evolution of the error in a convenient manner. Definitions for quasi-convexity and supporting results are presented in the appendix.

\begin{lem} If $\mathcal{U}^N$ is a symmetric transmission policy, then the error evolves according to   \begin{eqnarray} \label{eq:Error} E_{n+1} &=& \begin{cases} a E_n   + W_n & \text{if } L_{n+1} = 0 \\ 0 & \text{if } L_{n+1} =1. \end{cases}\end{eqnarray}\end{lem}

\begin{pf} This is in principle equivalent to \cite[Proposition 3.1]{GabrielAllerton}. The difference is that here $\mathcal{U}^N$ is a symmetric policy; not a symmetric threshold policy as in  \cite[Proposition 3.1]{GabrielAllerton}. However, the proof only relies on the symmetric nature of the policy. \end{pf}

The convenient form of the error evolution in (\ref{eq:Error}) is possible due to the symmetric assumption. For symmetric policies, when $L_n=0$ the optimal estimate $\hat{X}_n$ is the same whether a transmission was attempted or not. The remote estimator's belief $f_{X_n|V^n,R^n}$ depends on the value of $R_n$; however, its mean, which is the optimal estimate, does not.

The problem can be considered a Markov decision process with state $(E_{n-1},Q_{n})$, input $R_n$, and noise $(W_{n-1}, C_n)$. The cost to be minimized is  \begin{eqnarray*} \sum_{n=1}^N E[E_n^2]. \end{eqnarray*}  The associated dynamic programming recursion is given by \begin{eqnarray*}  V_{N+1}(e_{N},q_{N+1}) &=& e_N^2, \end{eqnarray*}  \begin{align} \label{symDP} V_n(e_{n-1},q_n)  &= \min\{ C^0_n(e_{n-1},q_n), C^1_n (e_{n-1},q_n) \}  , \end{align} for $n=1,\dots N$ with \begin{eqnarray*} C^0_n(e, q) &\eqdef& e^2+E_{W}[V_{n+1}(ae+W, q^0)] \\ C^1_n(e, q) &\eqdef& p_{q}e^2 +  p_{q} E_{W}[V_{n+1}(ae+W, q^1)] \\& & \qquad  + (1-p_{q}) E_W[V_{n+1}(W,q^1)] , \end{eqnarray*} and $q^0\eqdef \mathcal{M}^s(q,0)$, $q^1\eqdef \mathcal{M}^s(q,1)$, $p_{q} \eqdef \mathcal{M}^o(q)$ and $W$ distributed $\mathcal{N}(0,\sigma^2)$. 

\begin{lem} \label{lemma:Qconvex} For $n\in \{1,\dots ,\, N+1\}$ and $q\in \mathbb{Q}$, the cost-to-go functions $V_n(e_{n-1}, q)$ are quasi-convex and symmetric in $e_{n-1}$. The minimum value is $V_n(0,q)$. \end{lem}

\begin{pf}We show that $V_n(e_{n-1},q)$ is a symmetric and non-decreasing function in $|e_{n-1}|$. This implies $V_n(e_{n-1},q)$ is quasi-convex by Lemma \ref{lemma:StoC}. The proof is by induction. The claim holds for the initial case, $V_{N+1}(e_{N}, q_{N+1})=e_N^2$. Assume $V_{n+1}(e_{n},q_{n+1})$ is symmetric and non-decreasing in $|e_n|$.    $ V_n(e_{n-1},q_n) $ is the minimum between $C^0_n(e_{n-1},q_n)$ and $C^1_n(e_{n-1},q_n)$. By Lemma \ref{lemma:Expectation}, $ E_{W}[V_{n+1}(ae_{n-1}+W, q_n^i)] $ is symmetric and non-decreasing in $|e_{n-1}|$ for $i=0,1$.  $C^0_n(e_{n-1},q_n)$ and $C^1_n(e_{n-1},q_n)$ are symmetric and non-decreasing in $|e_{n-1}|$ because they are the sum of two such functions. Thus by Lemma \ref{lemma:Pmin}, $ V_n(e_{n-1},q_n) $ is symmetric and non-decreasing in $|e_{n-1}|$.  \end{pf}

\begin{thm}\label{thm:smallp}There exists a $v>0$ such that if for all $q\in \mathbb{Q}$ \begin{eqnarray*} p_q < \frac{1}{1+v},\end{eqnarray*} then the optimal symmetric transmission policy is a threshold policy.   \end{thm}

Several lemmata will be presented to aid in the proof of this theorem. Let $h^q_n(e) \eqdef E_W[V_n(ae+W,q)]$. Also define, \begin{eqnarray*} \mathcal{O}_n(e,q) \eqdef \lim_{\delta \downarrow 0} \frac{h_n^{q}(e+\delta) - h^{q}_n(e)}{(e+\delta)^2-e^2}.\end{eqnarray*}

\begin{lem} \label{generalLemma}  For $e\geq 0$ and $q\in \mathbb{Q}$, if \begin{eqnarray} \label{genLemAss} p_q \mathcal{O}_{n+1}(e,q^1) < (1-p_q) +\mathcal{O}_{n+1}(e, q^0),  \end{eqnarray}  then the optimal symmetric transmission policy for stage $n$ is a threshold policy. \end{lem}

\begin{pf} We show that if (\ref{genLemAss}) holds,  any non-threshold, symmetric policy is not the optimal symmetric transmission policy. 

For a non-threshold, symmetric policy $\mathcal{S}_n$ there exists a $q\in \mathbb{Q}$ and $c\geq0$ such that $\mathcal{S}_n(c,q) = 1$ but $\mathcal{S}_n(c+\delta,q) = 0$ for small $\delta>0$. Since $\mathcal{S}_n(c,q) = 1$ from (\ref{symDP}) we have $C_n^0(c,q)\geq C_n^1(c,q)$. Also, since $\mathcal{S}_n(c+\delta,q) = 0$ we have  $C_n^0(c+\delta,q) \leq C_n^1(c+\delta,q)$. By subtracting these equations we have $C_n^0(c+\delta,q)- C_n^0(c,q) \leq C_n^1(c+\delta,q) -  C_n^1(c,q)$. By rearranging terms this becomes \begin{eqnarray*} p_q[h_{n+1}^{q^1}(c+\delta) - h^{q^1}_{n+1}(c)] \geq (1-p_q) [(c+\delta)^2-c^2] \\ \ \ + h_{n+1}^{q^0}(c+\delta) - h_{n+1}^{q^0}(c). \end{eqnarray*} Dividing by $(c+\delta)^2 - c^2$ and taking the limit $\delta \downarrow 0$ yields \begin{eqnarray} \label{eq:toContradict} p_q \mathcal{O}_{n+1}(c,q^1) \geq (1-p_q) +\mathcal{O}_{n+1}(c, q^0).   \end{eqnarray} Contradicting the assumption. Thus, the optimal policy is a threshold policy. \end{pf}

\begin{remark} The condition in Lemma \ref{generalLemma}, garuntees that $C_n^0$ increases more than $C_n^1$ at every estimation error $e$. Clearly, this is a condition that leads to threshold transmission policies.  \end{remark}

 \begin{lem} \label{curve} For all $e\in \mathbb{R}$ and $q\in \mathbb{Q}$, \begin{eqnarray*}  \mathcal{O}_n(e,q) \leq v_n', \end{eqnarray*} with $v_n' \eqdef 2a^2(N+1-n)+a^2$. \end{lem} 

\begin{pf} We show inductively that for all $e\in \mathbb{R}$ and $q \in \mathbb{Q}$, there exists a $ v_n'$ such that $\mathcal{O}_{n+i}(e,q) \leq v_n' $, for $i=1 \, \dots \, (N+1-n)$.

This property holds for $N+1$, since $h_{N+1}^q =a^2e^2+\sigma^2$ and  $\mathcal{O}_{N+1}(e,q)=a^2$. Thus, $v_{N+1}'=a^2$.

Assume the property holds for $n+1$ with $v_{n+1}'$. We will show the property holds for $n$. For a specific  $e$ and $\omega$, there are two cases $V_n(ae+\omega,q) =C^0$ or $V_n(ae+\omega,q)=C^1$, see (\ref{symDP}). We prove the statement for the case when $V_n(ae+\omega,q) =C^0$. The other case yields the same result and is analogous.

\begin{figure}[t]
\vspace{7pt}
\centering
\includegraphics[width=0.5\textwidth]{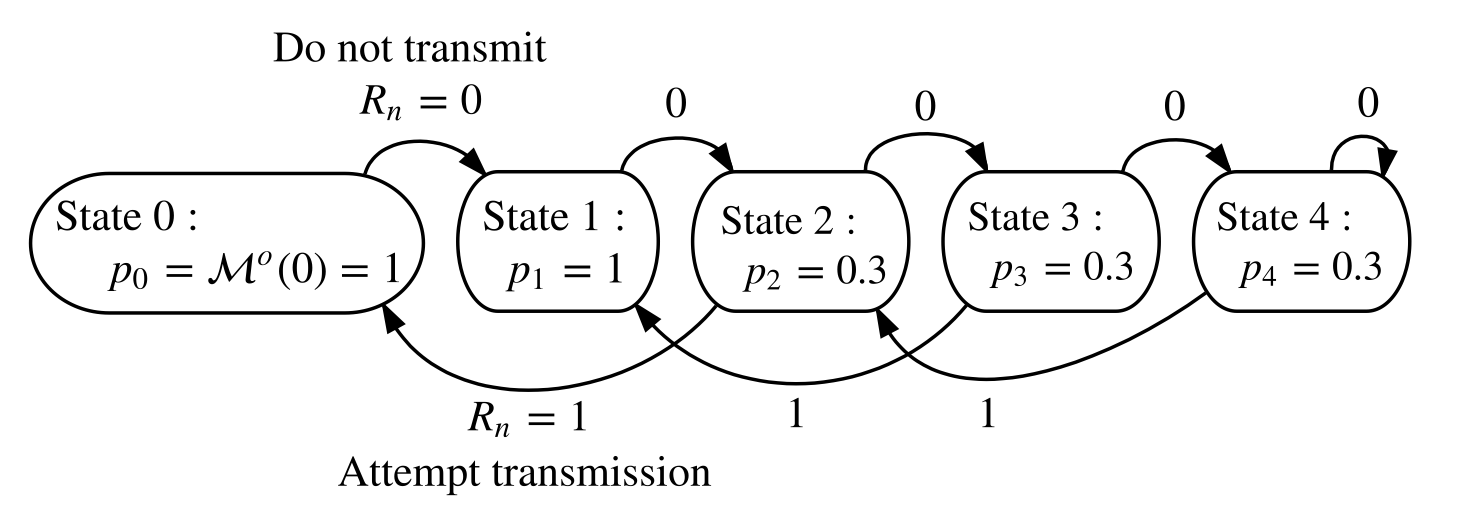}
\caption{ FSM model for an energy harvesting channel. State $i$ represents the energy currently stored in the battery.  The arcs represent channel state transitions which depend on whether a transmission is attempted. }
\label{fig:harvest}
\end{figure}

Equation (\ref{big2}), on the previous page, is obtained for the case using $V_n(ae+\omega,q) =C^0$ and using the bound \begin{eqnarray*}V_n(a(x+\delta)+\omega,q)&\leq &(a(x+\delta)+\omega)^2 \\ && + \,\,E[V_{n+1}(a(x+\delta)+\omega, q^1 )].\end{eqnarray*} The right hand side of (\ref{big2}) is comprised of two terms. The first term is upper bounded by a $2a^2 + \frac{a\omega}{x}$. 

Next, we take the expectation of (\ref{Big3}) with respect to $\omega$ and then the limit with respect to $\delta$. Using the inductive hypothesis to bound the second term by $v_{n+1}'$, this yields \begin{eqnarray*} \mathcal{O}_n(e,q) \leq 2a^2 + v_{n+1}'. \end{eqnarray*} Thus, with $v_n' = 2a^2+v_{n+1}'$ the induction is complete. We see that for all $n$, $v_n' = 2a^2(N+1-n) + a^2$ is an adequate bound.
  \end{pf}

\begin{pf}[of Theorem \ref{thm:smallp}] Using the bound $v = v_1'$ from Lemma \ref{curve}, we proceed by contradiction. We show that any non-threshold, symmetric transmission policy violates the assumption $p_q < \frac{1}{1+v}$.

Following identical arguments as in Lemma \ref{generalLemma}, we have from (\ref{eq:toContradict}) \begin{eqnarray*}  p_q \mathcal{O}_n(e,q^1) &\geq& (1-p_q) +\mathcal{O}_n(e, q^0) \\ &\geq& 1-p_q,\end{eqnarray*} since $\mathcal{O}_n(e,q)\geq 0$ by Lemma \ref{lemma:Qconvex}. Rearanging and using the bound on $\mathcal{O}_n(e,q^1)$ gives \begin{eqnarray*} p_q \geq \frac{1}{1+\mathcal{O}_n(e,q^1)} \geq \frac{1}{1+v}. \end{eqnarray*} Contradicting the assumption. Thus, the optimal policy is a threshold policy.

 \end{pf}


\section{Applications} \label{sec:appsBig}

\subsection{Energy harvesting channel application} \label{sec:apps}

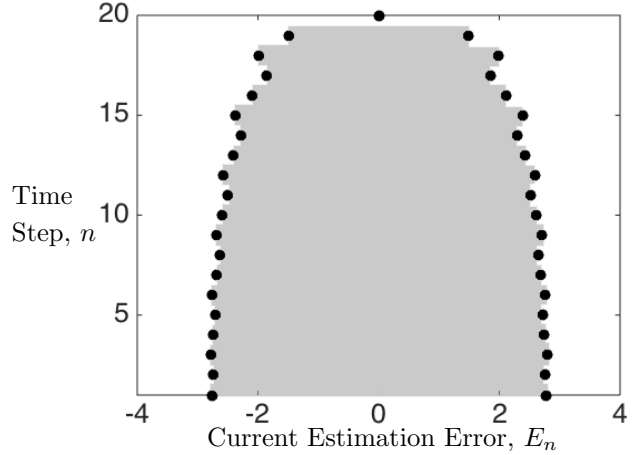
\begin{figure}[t]
\centering
\begin{tikzpicture}[xscale=1, yscale=1]
\node at (0,0) {\includegraphics[width=0.45\textwidth]{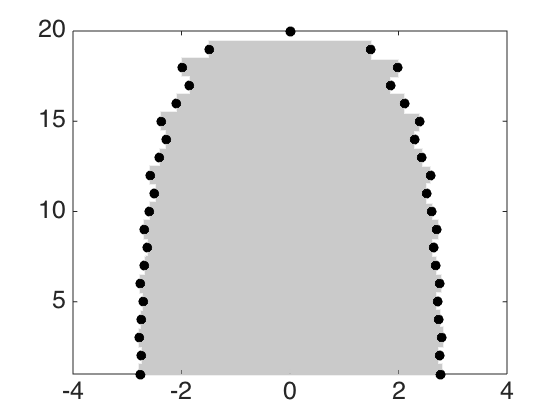}};
\node[above right] at (-4.8,0){Time};
\node[below right] at (-4.8,0){Step, $n$};
\node at (.2,-3){Current Estimation Error, $E_n$};
\end{tikzpicture}
\caption{Optimal symmetric transmission policy while in channel state $2$ of the use-dependent packet-drop channel as shown in figure \ref{fig:harvest}. This transmission policy was calculated using the values $a=1.1$, $\sigma=1$,  and $N=20$ by approximating the value functions. For errors less than the left black dots a transmission is attempted. For errors greater than the right black dots a transmission is attempted. Inside the gray region, no transmission is sent. } 
  \label{fig:taus20}
\end{figure}

A model of a wireless communication channel with energy harvesting capabilities is presented in this section. This channel is modeled with a use-dependent packet-drop channel. Many different problem formulations addressing remote estimation over a battery powered channel have been considered: see \cite{ulukus2015energy}, \cite{OzelDeterministic} and \cite{harvestNayyar} and the references therein.  

Consider a battery operated channel with a capacity of 4 energy units. Assume energy is harvested deterministically, as in \cite{OzelDeterministic},  at $1$ energy unit per time step. Transmitting requires $2$ units of energy and no energy is harvested during transmission. At each time step, the decision of whether to transmit is made. 

To model the battery dynamics, the FSM shown in figure \ref{fig:harvest} is used. The channel states are $\mathbb{Q}=\{0,1,2,3,4\}$. Channel state $q$ denotes that the battery has $q$ energy units. If a transmission is attempted $R_n=1$, then the battery level is reduced by $2$ energy units. Thus the channel state state $q$ transitions to state $q-2$. If a transmission is not attempted $R_n=0$, then the battery level increases by $1$ as long as the battery is not already at capacity. Thus, the channel transitions from state $q$ to state $\min \{q+1,4\}$. Obviously, in states $0$ and $1$ transmitting is not allowed due to insufficient energy. 

The probability of drop for each state capable of transmitting is $0.3$. Transmission is not possible in states $0$ and $1$ but we assign a drop probability of $1$ for consistency.

This energy harvesting channel is clearly a use-dependent packet-drop channel. We assume that the encoder receives acknowledgements of the transmissions and that the remote estimator can distinguish between a drop and no transmission attempt. Interestingly, from Theorem \ref{thm:thres} we have that the optimal transmission policy may not be symmetric in the estimation error even though the cost is symmetric in the estimation error and the noise is zero-mean and symmetric.

\subsubsection{Numerical example} \label{sec:num}

We numerically calculated the optimal symmetric transmission policies for this example when $a=1.1$, $\sigma=1$ and $N=20$. The optimal symmetric transmission policy for channel state $2$ is shown in figure \ref{fig:taus20}.

Notice that the optimal symmetric transmission policy is a threshold policy, even though the conditions of Theorem \ref{thm:smallp} are not satisfied. In fact, every $p_2,\, p_3,\,p_4 \in [0,1]$ that we tested has an optimal symmetric transmission policy that is a threshold policy. This suggests that for these channel dynamics, threshold transmission policies are optimal among all symmetric strategies.

In Theorem \ref{thm:smallp}, no assumptions were made about the size of the channel state space or the channel state dynamics. For specific channel dynamics  or classes of channel dynamics weakening the condition in Theorem \ref{thm:smallp} may be possible.

\subsection{Operator task shedding} \label{app:Operator}

In this section, we seek to optimize a decision support system for human operators tracking a dynamic target. 

\begin{figure}[t] 
\vspace{7pt}
\centering
\includegraphics[width=0.2\textwidth]{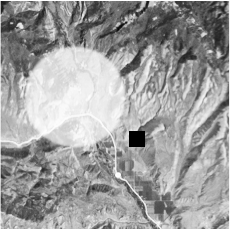}
\caption{Example of a display presented to an operator for the task shedding application. The target is the black square. The visual search task consists of the operator identifying if the target is inside the white region and optionally logging its location if it is outside the region. We seek to design the white regions dynamically to help the operator manage their time. } 
  \label{fig:mask}
  \end{figure}

Consider a human operator managing multiple UAVs. Tracking a dynamic target is one of operator's many tasks. A video feed is presented to the operator (see figure \ref{fig:mask} for an example of the video feed). The white region is drawn on the video feed by the decision support system. The operator's task is to indicate if the target is inside this region. If outside the region the operator is requested to log the target's current location; however, the operator is allowed to not log the target's location if other tasks seem more vital. \cite{slefAdaptive} perform experiments in a similar setting.

We seek to dynamically optimize the white regions in order to help the operator manage their time appropriately. If the regions are large, the target's location is not well known. If the regions are small, then the target's location is frequently requested. This increases the operator's workload and the likelihood the operator will ignore the request. The channel state is used to model operator workload. The optimal transmission policies define the optimal white regions and manage the tradeoff between accuracy and workload.

Yerkes-Dodson's law quantifies the tradeoff between operator performance and workload, see \cite{yerkes1908relation}.  Yerkes-Dodson's law states that the operator performs poorly if the workload is very high or very low. Optimizing operator decision support systems using Yerkes-Dodson's law as an operator model is also investigated in  \cite{savla} and \cite{srivastava}. In \cite{savla}, the workload impacts the time to complete tasks such that under high workload situations the operator completes tasks slowly. The authors find optimal policies specifying when to present the operator with tasks in order to maximize throughput.  In \cite{srivastava}, not all tasks must be completed and the questions of which tasks to assign, for how long, and with how much rest in-between are addressed.

In contrast to \cite{savla} and \cite{srivastava} and motivated by \cite{schulte}, we assume that the operator workload impacts the likelihood that the operator will ignore a request for information.

\begin{figure}[t]
\vspace{7pt}
\centering
\includegraphics[width=0.5\textwidth]{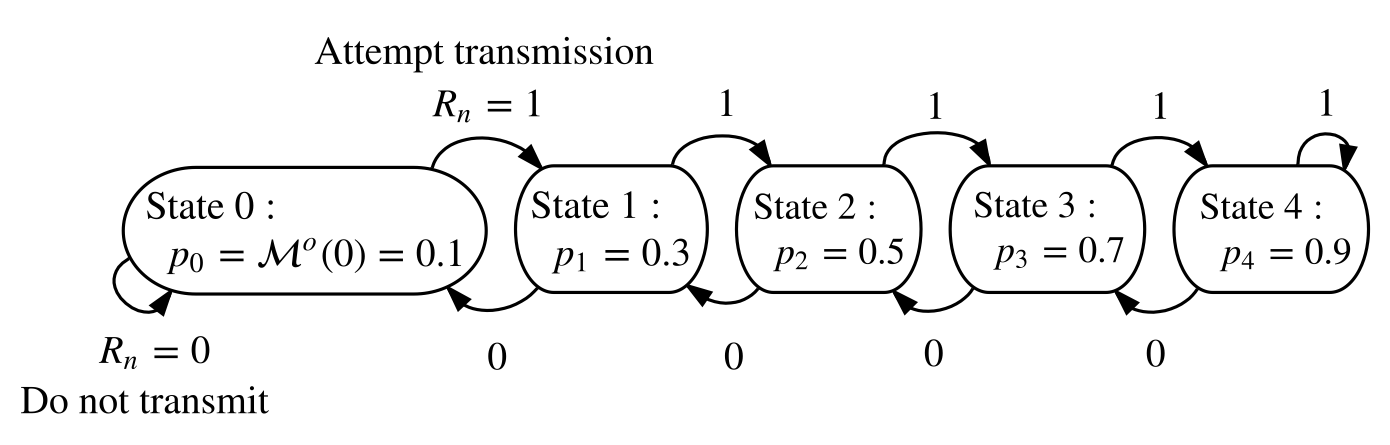}
\caption{ FSM model for human operator workload. The workload is a function of the average number of requests over the last $4$ time steps. State $i$ represents that $i$ requests have occurred in that last $4$ time steps. The arcs represent transitions of the channel state which depend on whether a transmission is requested.  }
\label{fig:chain}
\end{figure}

We consider the operator's workload a function of the average number of requests over the last $k$ time steps, \begin{eqnarray*} \frac{1}{k} \sum_{i=n-k}^n r_i .\end{eqnarray*} If the average is high, the operator is prone to shed tasks. This workload model has memory and can be envisioned as the finite state machine in figure \ref{fig:chain}. State $q$ represents $q$ requests occurring in the last $k$ steps.

To formulate this as a use-dependent packet-drop channel we take the target's location to be the system state, $X_n$. The target being outside the white region represents an attempted transmission $R_n=1$. The transmission policy $\mathcal{U}_n$ defines the  white region. 

We have modeled this application as a use-dependent packet-drop channel. By Theorem \ref{thm:smallp} if the operator is unlikely to ignore requests,  $p_n < 1/(1+v)$, then the optimal symmetric white regions are threshold policies. This is desirable since non threshold policies represent white regions that are not connected and may mislead operators.

The numerical example below suggests that threshold policies are the best symmetric policies even if the operator is likely to ignore requests. We believe this is due to the simple structure of the channel dynamics.

Note in this example $X_n$ is two dimensional; however, in our formulation $X_n$ is scalar. Under suitable independence assumptions, the results are applicable to higher dimensions.

\subsubsection{Numerical example} \label{sec:num}

We numerically find optimal symmetric transmission policies for this example when $a=1.1$, $\sigma=1$ and $N=20$. The channel dynamics and drop probabilities are shown in figure \ref{fig:chain}. The optimal symmetric transmission policies are calculated by approximating the value functions in (\ref{symDP}). In figure \ref{fig:taus20}, the optimal policies for channel sstates $0$ and $1$ are shown. It can be seen that the policies are symmetric. In fact, for all drop probabilities $p_0,\,p_1,\,p_2,\,p_3,\,p_4 \in [0,1]$ that were simulated, the optimal transmission policies were threshold policies.


\section{Conclusion}

We investigated optimal transmission policies for a remote estimation problem over a use-dependent packet-drop channel. We presented structural results for the optimal transmission policies under two different assumptions. Also, two examples were presented. An example application to energy harvesting channels and an example application to mixed initiative teams with human operator's performing visual search tasks were discussed.
 
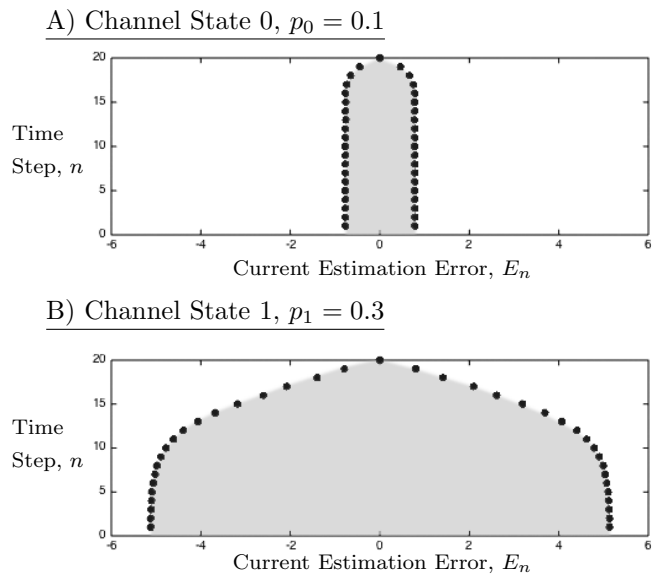
\begin{figure}[t]
\centering
\begin{tikzpicture}[xscale=1, yscale=1]
\node at (0,0) {\includegraphics[width=0.5\textwidth]{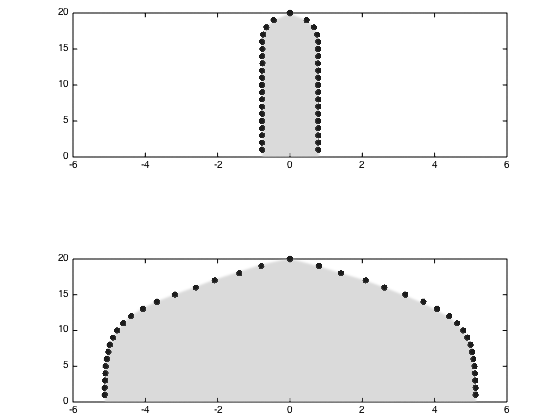}};
\node at (-2,-.2){\underline{B) Channel State $1,\, p_1 = 0.3$}};
\node at (-2,3.65){\underline{A) Channel State $0,\, p_0 = 0.1$}};
\node[above right] at (-4.8,-1.9){\small Time};
\node[below right] at (-4.8,-1.9){\small Step, $n$};
\node[above right] at (-4.8,2){\small Time};
\node[below right] at (-4.8,2){\small Step, $n$};
\node at (.2,-3.5){\small Current Estimation Error, $E_n$};
\node at (.2,.4){\small Current Estimation Error,  $E_n$};
\end{tikzpicture}
\caption{Optimal symmetric policies for the use-dependent packet-drop channel with dynamics as shown in figure \ref{fig:chain}. This transmission policy was calculated using the values $a=1.1$, $\sigma=1$,  and $N=20$. Part A of the figure plots the transmission policy for channel state $0$. Part B plots the policy for channel state $1$. For errors less than the left black dots a transmission is attempted. For errors greater than the right black dots a transmission is attempted. Inside the gray region, no transmission is sent. } 
  \label{fig:taus20}
\end{figure}

\bibliography{refs}             

\begin{thebibliography}{16}
\providecommand{\natexlab}[1]{#1}
\providecommand{\url}[1]{\texttt{#1}}
\providecommand{\urlprefix}{URL }
\expandafter\ifx\csname urlstyle\endcsname\relax
  \providecommand{\doi}[1]{doi:\discretionary{}{}{}#1}\else
  \providecommand{\doi}{doi:\discretionary{}{}{}\begingroup
  \urlstyle{rm}\Url}\fi

\bibitem[{Athans(1972)}]{Athans}
Athans, M. (1972).
\newblock On the determination of optimal costly measurement strategies for
  linear stochastic systems.
\newblock \emph{Automatica}, 8(4), 397 -- 412.

\bibitem[{Hajek et~al.(2008)Hajek, Mitzel, and Yang}]{paging}
Hajek, B., Mitzel, K., and Yang, S. (2008).
\newblock Paging and registration in cellular networks: Jointly optimal
  policies and an iterative algorithm.
\newblock \emph{Information Theory, IEEE Transactions on}, 54(2), 608--622.

\bibitem[{Lipsa and Martins(2009)}]{GabrielAllerton}
Lipsa, G.M. and Martins, N.C. (2009).
\newblock Optimal state estimation in the presence of communication costs and
  packet drops.
\newblock In \emph{Proceedings of the 47th Annual Allerton Conference on
  Communication, Control, and Computing}, Allerton'09, 160--169.

\bibitem[{Lipsa and Martins(2011)}]{GabrielTrans}
Lipsa, G. and Martins, N. (2011).
\newblock Remote state estimation with communication costs for first-order lti
  systems.
\newblock \emph{Automatic Control, IEEE Transactions on}, 56(9), 2013--2025.

\bibitem[{Nayyar et~al.(2012)Nayyar, Basar, Teneketzis, and
  Veeravalli}]{harvestNayyar}
Nayyar, A., Basar, T., Teneketzis, D., and Veeravalli, V.V. (2012).
\newblock Optimal strategies for communication and remote estimation with an
  energy harvesting sensor.
\newblock \emph{CoRR}, abs/1205.6018.

\bibitem[{Ozel et~al.(2011)Ozel, Yang, and Ulukus}]{OzelDeterministic}
Ozel, O., Yang, J., and Ulukus, S. (2011).
\newblock Optimal scheduling over fading broadcast channels with an energy
  harvesting transmitter.
\newblock In \emph{Computational Advances in Multi-Sensor Adaptive Processing
  (CAMSAP), 2011 4th IEEE International Workshop on}, 193--196.

\bibitem[{Savla and Frazzoli(2012)}]{savla}
Savla, K. and Frazzoli, E. (2012).
\newblock A dynamical queue approach to intelligent task management for human
  operators.
\newblock \emph{Proceedings of the IEEE}, 100(3), 672--686.

\bibitem[{Schulte and Donath(2011{\natexlab{a}})}]{slefAdaptive}
Schulte, A. and Donath, D. (2011{\natexlab{a}}).
\newblock Measuring self-adaptive uav operators' load-shedding strategies under
  high workload.
\newblock In \emph{Proceedings of the 9th International Conference on
  Engineering Psychology and Cognitive Ergonomics}, EPCE'11, 342--351.

\bibitem[{Schulte and Donath(2011{\natexlab{b}})}]{schulte}
Schulte, A. and Donath, D. (2011{\natexlab{b}}).
\newblock Measuring self-adaptive uav operators' load-shedding strategies under
  high workload.
\newblock In \emph{HCI (21)'11}, 342--351.

\bibitem[{Shamaiah et~al.(2010)Shamaiah, Banerjee, and Vikalo}]{submodular}
Shamaiah, M., Banerjee, S., and Vikalo, H. (2010).
\newblock Greedy sensor selection: Leveraging submodularity.
\newblock In \emph{Decision and Control (CDC), 2010 49th IEEE Conference on},
  2572--2577.

\bibitem[{Sinopoli et~al.(2004)Sinopoli, Schenato, Franceschetti, Poolla,
  Jordan, and Sastry}]{sinopoli}
Sinopoli, B., Schenato, L., Franceschetti, M., Poolla, K., Jordan, M., and
  Sastry, S. (2004).
\newblock Kalman filtering with intermittent observations.
\newblock \emph{Automatic Control, IEEE Transactions on}, 49(9), 1453--1464.

\bibitem[{Srivastava et~al.(2012)Srivastava, Surana, and Bullo}]{srivastava}
Srivastava, V., Surana, A., and Bullo, F. (2012).
\newblock Adaptive attention allocation in human-robot systems.
\newblock In \emph{American Control Conference (ACC), 2012}, 2767--2774.

\bibitem[{Ulukus et~al.(2015)Ulukus, Yener, Erkip, Simeone, Zorzi, Grover, and
  Huang}]{ulukus2015energy}
Ulukus, S., Yener, A., Erkip, E., Simeone, O., Zorzi, M., Grover, P., and
  Huang, K. (2015).
\newblock Energy harvesting wireless communications: A review of recent
  advances.
\newblock \emph{Selected Areas in Communications, IEEE Journal on}, 33(3),
  360--381.

\bibitem[{Vasconcelos and Martins(2013)}]{MarcosAllerton}
Vasconcelos, M. and Martins, N. (2013).
\newblock Estimation over the collision channel: Structural results.
\newblock In \emph{Communication, Control, and Computing (Allerton), 2013 51st
  Annual Allerton Conference on}, 1114--1119.

\bibitem[{Weissman(2010)}]{weissman2010capacity}
Weissman, T. (2010).
\newblock Capacity of channels with action-dependent states.
\newblock \emph{Information Theory, IEEE Transactions on}, 56(11), 5396--5411.

\bibitem[{Yerkes and Dodson(1908)}]{yerkes1908relation}
Yerkes, R.M. and Dodson, J.D. (1908).
\newblock The relation of strength of stimulus to rapidity of habit-formation.
\newblock \emph{Journal of comparative neurology and psychology}, 18(5),
  459--482.

\end{thebibliography}
                                                   







\appendix
\section{Quasi-Convex Functions}

In this appendix, definitions and results related to quasi-convex functions are presented.

\begin{define} A function $f:\mathbb{R}\rightarrow \mathbb{R}$ is quasi-convex if for $x,y\in \mathbb{R}$ and $\lambda \in [0,1]$  \begin{eqnarray*} f(\lambda x + (1-\lambda)y) \leq \max \{f(x), f(y) \}.   \end{eqnarray*} \end{define}

\begin{define} A function $f:\mathbb{R}\rightarrow \mathbb{R}$ is symmetric and non-decreasing in $|x|$ if for $0\leq x<y$,   \begin{eqnarray*} f(x)  &=& f(-x) \text{  and} \\ f(x) &\leq& f(y).    \end{eqnarray*} \end{define}

\begin{lem} \label{lemma:StoC} If $f:\mathbb{R}\rightarrow \mathbb{R}$ is symmetric and non-decreasing in $|x|$ then $f$ is quasi-convex.  \end{lem}

\begin{pf} For $x,y\in \mathbb{R}$, without loss of generality let $|y|>|x|$. Note $f(y)\geq f(x)$. For $\lambda\in[0,1]$, since $|\lambda x + (1-\lambda) y| \leq |y|$, we have $f(\lambda x + (1-\lambda) y) \leq f(y)$.  \end{pf}

\begin{lem}\label{lemma:Pmin} Let $f,g$ be symmetric and non-decreasing in $|x|$. The function $h(x) = \min\{f(x), g(x) \}$ is symmetric and non-decreasing in $|x|$. \end{lem}

\begin{pf} First, we show $h$ is symmetric. For $x\in\mathbb{R}$, \begin{eqnarray*} h(-x) &=& \min\{ f(-x), g(-x) \}\\  &=& \min\{ f(x), g(x) \} \\ &=&h(x). \end{eqnarray*} 

We now show $h$ is non-decreasing. For $0\leq x < y$, \begin{eqnarray*} h(x) &=& \min\{ f(x), g(x) \}\\  &\leq& \min\{ f(y), g(y) \} \\ &=&h(y). \end{eqnarray*} \end{pf}

\begin{lem} \label{lemma:Expectation} Let $f$ be a symmetric and non-decreasing in $|x|$, $W$ a random variable distributed $\mathcal{N}(0,\sigma^2)$ and $a\in \mathbb{R}$. The function $h(x) = E_W[f(ax+W)]$ is symmetric and non-decreasing in $|x|$. \end{lem}

\begin{pf} First, we show $h$ is symmetric. For $x\in\mathbb{R}$, \begin{eqnarray*} h(-x) &=& \int_{- \infty}^{\infty} f(-ax+w) \eta e^{\frac{-w^2}{2\sigma^2}} dw  \\  &=& \int_{-\infty}^{\infty} f(-ax-w') \eta e^{\frac{-w'^2}{2\sigma^2}} dw' \\&=& h(x) \end{eqnarray*} where $\eta=\frac{1}{\sqrt{2\pi \sigma^2}}$. The second equality holds by change of variables $w' = w$.

We now show $h$ is non-decreasing. Let $0\leq x < y$. Using the symmetry of $f$, with $\eta \eqdef \frac{1}{\sqrt{2\pi\sigma^2}}$, $h(x)$ can be written, \begin{eqnarray*} h(x) =\int_{0}^\infty f(w)\eta [e^{\frac{-(w-ax)^2}{2\sigma^2}}+e^{\frac{-(-w-ax)^2}{2\sigma^2}}] dw. \end{eqnarray*} Consider \begin{eqnarray*} h(y)-h(x)  &=& \int_{0}^\infty f(w)\eta g(w) dw, \end{eqnarray*} with \begin{eqnarray*} g(w)\eqdef &&e^{\frac{-(w-ay)^2}{2\sigma^2}}+e^{\frac{-(-w-ay)^2}{2\sigma^2}} \\ && \qquad - \left[e^{\frac{-(w-ax)^2}{2\sigma^2}}+e^{\frac{-(-w-ax)^2}{2\sigma^2}}\right]. \end{eqnarray*} There exists a $\bar{w}>0$ such that $g(w)<0$ for $0<w< \bar{w}$ and $g(w)\geq0$ for $w\geq \bar{w}$. So \begin{flalign*} h(y)-h(x)  &= \int_{0}^{\bar{w}} f(w)\eta g(w) dw +\int_{\bar{w}}^\infty f(w)\eta g(w) dw \\ &\geq f(\bar{w})\int_{0}^{\bar{w}} \eta g(w) dw +f(\bar{w})\int_{\bar{w}}^\infty \eta g(w) dw \\ &= f(\bar{w}) [1 -1] = 0.\end{flalign*} Thus, $h(y)\geq h(x)$.

\end{pf}

\end{document}